\documentclass[11pt,draftcls,onecolumn]{IEEEtran}
\usepackage{etex} % pour compilation Mustapha
\usepackage{amsmath,graphicx,amssymb,epsfig,setspace,multicol,cite}
\usepackage{multirow}
\usepackage{float,dsfont}
\usepackage[latin1]{inputenc}
\usepackage{graphicx}
\usepackage{mathtools}
\usepackage{ulem} % pour barre un texte
\usepackage{bbm} % pour fonction indicatrice
\usepackage{schemabloc}  %% CE QU'IL FAUT
\usepackage{epsfig}       %% RAJOUTER POUR LES
\usepackage{pst-all}      %% DESSINS
\usepackage{cancel}

\usepackage{tikz}
\usepackage{epstopdf}
\usetikzlibrary{positioning}

\DeclareTextSymbol{\degre}{T1}{6}
\DeclareTextSymbol{\degre}{OT1}{23}

\newcommand\cov{{\rm Cov}}

\newcommand\ve{{\rm vec}}

\newcommand\e{{\rm E}}
\newcommand\pardef{ \stackrel{{\rm def}}{=} }

\newtheorem{result}{Result}

\def\QED{\mbox{\rule[0pt]{1.4ex}{1.4ex}}}

\usepackage{soul} % pour barrer un texte
\usepackage{color}

\begin{document}

\title{The Gaussian data assumption does not always lead to the largest Cram\'er-Rao bound}
\author{Jean-Pierre Delmas and Habti Abeida
}
\maketitle

%%%%%%%%%%%%%%%%%%%%%%%%%%%%%%%%%%%%%%%%%%%%%%%%ù
%%%%%%%%%%%%%%%%%%%%%%%%%%%%%%%%%%%%%%%%%%%%%%%%%%%%%%%%%%%%%%%%%\`{u}\`{u}
%
%\begin{abstract}
%x
%\end{abstract}
%%%%%%%%%%%%%%%%%%%%%%%%%%%%%%%%%%%%%%%%%%%%%%%%%%%%%%%%%%%%%%%%%%%%%%%%%%%%%%
%
%\begin{IEEEkeywords}
%Whittle's formula, Bangs's formula, Cram\'er-Rao bound, compound-Gaussian process, Elliptical symmetric distributions, stationary, purely cyclostationary, almost cyclostationary,  DOA estimation.
%\end{IEEEkeywords}

\maketitle
\centerline{
Published in Lecture Notes Column to IEEE SP Magazine (vol. 43 no. 5, pp. 42-46, September 2026)
}
\date{\today}
%

%%%%%%%%%%%%%%%%%%%%%%%%%%%%%%%%%%%%
%%%%%%%%%%%%%%%%%%%%%%%%%%%%%%%%%%%%%%%%%%%%%%%%%%%%%%%%%%%%%%%%%%%%%%%
\section{Scope}
\label{sec:Scope}
%%%%%%%%%%%%%%%%%%%%%%%%%%%%%%%%%%%%%%%%%%%%%%%%%%%%%%%%%%%%%%%%%%%%%%
%
This lecture note addresses the common misconception that the Gaussian distribution always yields the largest Cram\'er-Rao Bound (CRB). 
We show that this property only holds under restrictive conditions: specifically, when the mean and covariance parameters are decoupled in the Fisher Information Matrix (FIM), when the parameter of interest lies in the mean vector and when there are no additive nuisance parameters. Beyond this framework, we provide counterexamples demonstrating that non-Gaussian distributions can produce larger CRB.
%
%%%%%%%%%%%%%%%%%%%%%%%%%%%%%%%%%%%%%%%%%%%%%%%%%%%%%%%%%%%%%%%%%%%%%%
\section{Relevance}
\label{sec:Relevance}
%%%%%%%%%%%%%%%%%%%%%%%%%%%%%%%%%%%
%
The widespread use of the Gaussian distribution in statistical signal processing and many other applied fields  stems from its analytical tractability and support from the central limit theorem. Additionally, it is justified by the property that, among all distributions of data with fixed mean ${\bf m}$ and covariance matrix ${\bf \Sigma}$, it yields the largest CRB on the parameters.
This ensures that any optimum design based on the CRB under the Gaussian distribution of the data can be considered to be optimal in the sense of minimizing the largest CRB.
However, this critical third justification is often presented without the necessary caveats, giving rise to a widespread misconception. Standard references, such as \cite[Appendix B.3.26]{Stoica1997}, state this property without specifying the strict conditions under which it holds. While works like \cite{Stoica2011,Park2013,Stein2014} provide 
formal proofs, they are limited to the special case of a block-diagonal FIM for the mean and covariance parameters, and specifically when the parameter of interest is the mean. Their analyses do not address scenarios with a coupled FIM or when the parameter of interest lies in the covariance matrix, leaving an important gap that may cause confusion.

This lecture note rectifies this gap by introducing a comprehensive, general multivariate location-scale model. Our framework is designed to encompass the full scope of the problem: it allows parameters of interest in either the mean or the covariance and incorporates a necessary nuisance parameter to fully specify the distribution.   Consequently, this work clarifies a foundational concept in estimation theory. By providing a complete characterization of CRB maximization, this lecture note serves as an essential resource for both courses and advanced research in statistical signal processing, ensuring that future designs and analyses are built upon a correct theoretical foundation.
%
%%%%%%%%%%%%%%%%%%%%%%%%%%%%%%%%%%%%%%%%%%%%%%%%%%%%%%%%%%%%%%%%%%%%%%
\section{Prerequisites}
\label{sec:Prerequisites}
%%%%%%%%%%%%%%%%%%%%%%%%%%%%%%%%%%%
%
The reader is expected to have a basic background in probability and statistical signal processing. Some familiarity with the Slepian-Bangs formula, which provides a closed-form expression for the FIM in the Gaussian case (see the proofs in textbooks \cite[Th. 5.1]{Porat1994},\cite[Appendix B.3]{Stoica1997}), and with elliptically symmetric probability distributions may be helpful for a smoother reading. For complementary background on elliptically symmetric distributions, the reader may refer to \cite{Delmas2024}, although the presentation remains self-contained.
%
%%%%%%%%%%%%%%%%%%%%%%%%%%%%%%%%%%%%%%%%%%%%%%%%%%%%%%%%%%%%%%%%%%%%%%
%\section{Notations}
%\label{sec:Notation}
%%%%%%%%%%%%%%%%%%%%%%%%%%%%%%%%%%%%
%The following notations are used throughout this lecture note.
%Matrices
%and vectors are represented by bold upper case and bold lower case
%characters, respectively. Vectors are by default in column
%orientation, while the superscript $T$ stands for transpose.
%The derivative of a $m \times 1$ scalar or vector function with respect to a $q \times 1$ vector are written in $m \times q$ matrix form.
%$\e(.)$, $\cov(.)$ and $|.|$ are the expectation, covariance and determinant operators, respectively.
%${\bf I}_n$ is the identity matrix of dimension $n$.
%$\ve ({\bf A})$ denotes the
%``vectorization'' operator that turns a matrix ${\bf A}$ into a vector by
%stacking the columns of the matrix one below another.
%This vector is used
%in conjunction with the Kronecker product ${\bf A}\otimes{\bf B}$
%as the block matrix whose $(i,j)$ block element is $a_{i,j}{\bf B}$.
%$x=_d y$ means that the random variables $x$ and $y$ have the same distribution.
%The acronyms r.v., p.d.f., i.i.d., and i.f.f. are used to represent "random variable", "probability density function",  "independent identically distributed", and  ''if and only if'', respectively.
%%%%%%%%%%%%%%%%%%%%%%%%%%%%%%%%%%%%%%%%%%%%%%%%%%%%%%%%%%%%%%%%%%%%%%
\section{Problem statement and solution}
\label{sec:Problem statement and solution}
%%%%%%%%%%%%%%%%%%%%%%%%%%%%%%%%%%%
%
\subsection{Problem statement and parameterization}
\label{sec:problem statement and parameterization}
We consider an observed random vector ${\bf x} = (x_{1},\ldots,x_{n})^{T}$ whose joint p.d.f. depends on two sets of parameters $\boldsymbol{\theta}_{1} \in \mathbb{R}^{q_{1}}$ and $\boldsymbol{\theta}_{2} \in \mathbb{R}^{q_{2}}$. These parameters determine, through differentiable and identifiable mappings\footnote{Identifiability means that ${\bf m}(\boldsymbol{\theta}'_{1})={\bf m}(\boldsymbol{\theta}''_{1}) \Rightarrow \boldsymbol{\theta}'_{1}=\boldsymbol{\theta}''_{1}$ and ${\bf \Sigma}(\boldsymbol{\theta}'_{2})={\bf \Sigma}(\boldsymbol{\theta}''_{2}) \Rightarrow \boldsymbol{\theta}'_{2}=\boldsymbol{\theta}''_{2}$.}, the mean vector ${\bf m}(\boldsymbol{\theta}_{1})$ and the positive definite covariance matrix ${\bf \Sigma}(\boldsymbol{\theta}_{2})$, respectively.  Depending on the estimation objective, one of these parameter vectors plays the role of the parameter of interest while the other is treated as a nuisance parameter\footnote{When $\boldsymbol{\theta}_{2}$ is the parameter of interest, the observation model is typically assumed to be zero-mean.}. In some applications, an additional nuisance parameter $\boldsymbol{\theta}_{3} \in \mathbb{R}^{q_{3}}$ may also be present.

We note that this general location-scale model encompasses both i.i.d.\ and non-i.i.d.\ random observations, thereby covering essentially all applications in statistical signal processing. By introducing the associated normalized random vector
${\bf y} \pardef {\bf \Sigma}^{-1/2}({\bf x}-{\bf m})$,
with p.d.f. $p_y({\bf y})$ that may depend on an additional parameter $\boldsymbol{\theta}_3$, the p.d.f. of ${\bf x}$ can be expressed in the location-scale form:
\begin{equation}
\label{eq:pdf general}
p_x({\bf x}) = |{\bf \Sigma}|^{-1/2} \, p_y\big({\bf \Sigma}^{-1/2}({\bf x}-{\bf m})\big).
\end{equation}
Let $\boldsymbol{\theta}_i$, with $i=1$ or $2$, denotes the parameter of interest. 
The CRB on $\boldsymbol{\theta}_i$ is derived from the FIM of
$\boldsymbol{\theta} \triangleq (\boldsymbol{\theta}_1^T, \boldsymbol{\theta}_2^T, \boldsymbol{\theta}_3^T)^T$,
denoted \({\bf I}(\boldsymbol{\theta})\), by
${\rm CRB}(\boldsymbol{\theta}_i) = [{\bf I}^{-1}(\boldsymbol{\theta})]_{(i,i)}$.

The next section presents the explicit derivation of this FIM for the general location-scale model. This is
directly motivated by the central question addressed in this lecture note: is whether the CRB on 
$\boldsymbol{\theta}_i$ under the Gaussian assumption is always greater than or equal to the CRB obtained 
under any other distribution sharing the same mean ${\bf m}$ and covariance matrix ${\bf \Sigma}$?
%
%%%%%%%%%%%%%%%%%%%%%%%%%%%%%%%%%%%%%%%%%%%%%%%%%%%%%%%%%%%%%%%%%%%%%%
\subsection{Fisher Information Matrix}
\label{sec:Fisher Information matrix}
%%%%%%%%%%%%%%%%%%%%%%%%%%%
%
The FIM associated with the parameter vector 
$\boldsymbol{\theta}$ 
is defined as
${\bf I}(\boldsymbol{\theta})
=\e[s_x(\boldsymbol{\theta})\, s_x^T(\boldsymbol{\theta})]$,
where $s_x(\boldsymbol{\theta}) = \left(\frac{\partial \ln p_x({\bf x})}{\partial \boldsymbol{\theta}}\right)^T$
denotes the score vector. Its $(i,j)$ block submatrix is therefore given by
$[{\bf I}(\boldsymbol{\theta})]_{(i,j)}
= \e[s_x(\boldsymbol{\theta}_i)\, s_x^T(\boldsymbol{\theta}_j)]$, $i,j=1,2,3$,
For the location-scale density in \eqref{eq:pdf general}, the score vectors 
$s_x(\boldsymbol{\theta}_1)$ and $s_x(\boldsymbol{\theta}_2)$ are obtained using 
the chain rule and the identity\footnote{where $\ve ({\bf A})$ denotes the
``vectorization'' operator that turns a matrix ${\bf A}$ into a vector by
stacking the columns of the matrix one below another and
where the Kronecker product ${\bf A}\otimes{\bf B}$
is the block matrix whose $(i,j)$ block element is $a_{i,j}{\bf B}$.}
${\bf \Sigma}^{-1/2}({\bf x}-{\bf m})
= [({\bf x}-{\bf m})^T \otimes {\bf I}_n] \, \ve({\bf \Sigma}^{-1/2})$,
which yields
\begin{eqnarray}
\label{eq:score theta1}
s_x(\boldsymbol{\theta}_1)
&=&
-\left(\frac{d{\bf m}}{d\boldsymbol{\theta}_1}\right)^T{\bf \Sigma}^{-1/2}\boldsymbol{\xi}({\bf y})
\\
\nonumber
s_x(\boldsymbol{\theta}_2)
&=&
-\frac{1}{2|{\bf \Sigma}|}\left(\frac{d|{\bf \Sigma}|}{d \boldsymbol{\theta}_2}\right)^T
\\
\label{eq:score theta2}
&+&
\left(\frac{d \ve({\bf \Sigma}^{-1/2})}{d\boldsymbol{\theta}_2}\right)^T
[{\bf \Sigma}^{-1/2}{\bf y}\otimes {\bf I}_n] \boldsymbol{\xi}({\bf y})
\end{eqnarray}
with\footnote{The derivative of a $m \times 1$ scalar or vector function with respect to a $q \times 1$ vector are written in $m \times q$ matrix form.} 
$\boldsymbol{\xi}({\bf y})\pardef \frac{1}{p_y({\bf y})}\left(\frac{d p_y({\bf y})}{d{\bf y}}\right)^T$.

In most applications, the p.d.f.\ $p_x({\bf x})$ is symmetric about the mean ${\bf m}$, which is equivalent to 
$p_y({\bf y})$ being an even function. In this case, from \eqref{eq:score theta1} and \eqref{eq:score theta2}, 
the score $s_x(\boldsymbol{\theta}_1)$ is an even function of ${\bf y}$, whereas $s_x(\boldsymbol{\theta}_2)$ 
is an odd function of ${\bf y}$. Consequently,
$\e[s_x(\boldsymbol{\theta}_1)\,s_x^T(\boldsymbol{\theta}_2)]
=\int_{\mathbb{R}^n} s_x(\boldsymbol{\theta}_1)s_x^T(\boldsymbol{\theta}_2)p_y({\bf y})\,d{\bf y}
={\bf 0}$,
and therefore the parameters $\boldsymbol{\theta}_1$ and $\boldsymbol{\theta}_2$ are decoupled in the FIM. 
This symmetry-based decoupling has been repeatedly used in the literature (see, e.g., \cite{Stoica2011,Park2013}), 
although a complete proof appears difficult to locate in prior references.
In the absence of a nuisance parameter $\boldsymbol{\theta}_3$, the CRB expressions therefore simplify to
\begin{equation}
\label{eq:CRB theta1 theta2}
{\rm CRB}(\boldsymbol{\theta}_1)=\big([{\bf I}(\boldsymbol{\theta})]_{(1,1)}\big)^{-1},
\qquad
{\rm CRB}(\boldsymbol{\theta}_2)=\big([{\bf I}(\boldsymbol{\theta})]_{(2,2)}\big)^{-1}.
\end{equation}

To compare these bounds with those obtained under the Gaussian assumption, we recall the 
Slepian-Bangs formula, which provides closed-form expressions for the FIM when 
${\bf x} \sim \mathcal{N}({\bf m}(\boldsymbol{\theta}_1),{\bf \Sigma}(\boldsymbol{\theta}_2))$:
\begin{align}
\label{eq:FIM Gaussian 11}
[{\bf I}(\boldsymbol{\theta})]_{(1,1)}
&=
\left(\frac{\partial {\bf m}}{\partial \boldsymbol{\theta}_1}\right)^T
{\bf \Sigma}^{-1}
\frac{\partial {\bf m}}{\partial \boldsymbol{\theta}_1},
\\
\label{eq:FIM Gaussian 22}
[{\bf I}(\boldsymbol{\theta})]_{(2,2)}
&=
\frac{1}{2}
\left(\frac{\partial \ve({\bf \Sigma})}{\partial \boldsymbol{\theta}_2}\right)^T
\big({\bf \Sigma}^{-1} \otimes {\bf \Sigma}^{-1}\big)
\frac{\partial \ve({\bf \Sigma})}{\partial \boldsymbol{\theta}_2}.
\end{align}
As a simple illustration, for $n$ i.i.d.\ univariate Gaussian observations 
${\cal N}(m,\sigma^2)$ with parameters $\theta_1=m$ and $\theta_2=\sigma^2$, 
\eqref{eq:FIM Gaussian 11} and \eqref{eq:FIM Gaussian 22} reduce to the classical Fisher information values
$I(m)=\frac{n}{\sigma^2}$
and
$I(\sigma^2)=\frac{n}{2\sigma^4}$.

The symmetry of the p.d.f.  allows us to establish the following result.
\begin{result}
\label{re:Gaussian best}
For all distribution whose p.d.f. $p_x({\bf x})$ is symmetric about the mean ${\bf m}$
with no additive nuisance parameter $\boldsymbol{\theta}_3$, the 
 Gaussian distribution leads to the largest CRB on the parameters $\boldsymbol{\theta}_1$ that parameterizes the mean.
  In other words,
${\rm CRB}(\boldsymbol{\theta}_1)
\;\leq\;
{\rm CRB}_{\rm Gaus}(\boldsymbol{\theta}_1)$,
where $\leq$ denotes matrix inequality in the positive semi-definite sense
and where ${\rm CRB}_{\rm Gau}(\boldsymbol{\theta}_1)$ denotes the CRB on $\boldsymbol{\theta}_1$ under the Gaussian distribution of ${\bf x}$.
\end{result}

{\it Proof:} 
Starting from \eqref{eq:score theta1}, the $(1,1)$ block of the FIM is given by
\begin{equation}
\label{eq:FIM theta1}
[{\bf I}(\boldsymbol{\theta})]_{(1,1)}
=
\left(\frac{\partial{\bf m}}{\partial \boldsymbol{\theta}_1}\right)^{\!T}
{\bf \Sigma}^{-1/2}
\e[\boldsymbol{\xi}({\bf y})\boldsymbol{\xi}^{T}({\bf y})]
{\bf \Sigma}^{-1/2}
\left(\frac{\partial{\bf m}}{\partial \boldsymbol{\theta}_1}\right).
\end{equation}
Then for any vector ${\bf v}\in\mathbb{R}^n$, the Cauchy-Schwarz inequality yields
\begin{align}
\nonumber
\left[\int_{\mathbb{R}^n} 
{\bf v}^T{\bf y}\,\frac{\partial p_y({\bf y})}{\partial {\bf y}}
{\bf v}d{\bf y}
\right]^2
&=
\left[
\int_{\mathbb{R}^n}
\bigl({\bf v}^T{\bf y}\, p_y^{1/2}({\bf y})\bigr)
\left(\frac{1}{p_y^{1/2}({\bf y})}
\frac{\partial p_y({\bf y})}{\partial {\bf y}}{\bf v}\right)d{\bf y}
\right]^2
\\
\label{eq:Cauchy-Schwarz inequality}
&\le
\left(\int_{\mathbb{R}^n} {\bf v}^T{\bf y}\,{\bf y}^T{\bf v} \, p_y({\bf y})d{\bf y}\right)
\left(\int_{\mathbb{R}^n} 
\frac{1}{p_y({\bf y})}\left(\frac{\partial p_y({\bf y})}{\partial {\bf y}}{\bf v}\right)^2 d{\bf y}\right).
\end{align}
Since $\e[{\bf y}{\bf y}^T]={\bf I}_n$, we have
$\int_{\mathbb{R}^n} {\bf v}^T{\bf y}{\bf y}^T{\bf v} \, p_y({\bf y})\, d{\bf y}
= {\bf v}^T {\bf I}_n {\bf v}$.
Moreover,
$\int_{\mathbb{R}^n}
\frac{1}{p_y({\bf y})}\left(\frac{\partial p_y({\bf y})}{\partial {\bf y}}{\bf v}\right)^2 d{\bf y}
= {\bf v}^T\,\e[\boldsymbol{\xi}({\bf y})\boldsymbol{\xi}^{T}({\bf y})]{\bf v}$.
Then by integration over $\mathbb{R}^n$ of the equality:
$\frac{\partial [y_i p_y({\bf y})]}{\partial y_j}=y_i \frac{\partial  p_y({\bf y})}{\partial y_j}+\delta_{i,j}p_y({\bf y})$ and interchanging differentiation and integration
in the term $\frac{\partial [y_i p_y({\bf y})]}{\partial y_j}$ permitted under regularity conditions, we obtain
$0
=\frac{\partial}{\partial y_j} [\e(y_i)]
=
\frac{\partial}{\partial y_j} [\int_{\mathbb{R}^n}y_ip_y({\bf y})d{\bf y}]
=\int_{\mathbb{R}^n}y_i \frac{\partial  p_y({\bf y})}{\partial y_j}d{\bf y}+\delta_{i,j}$.
Collecting
the identities for all $i$, $j$ yields $
\int_{\mathbb{R}^n}
{\bf y}\, \frac{\partial p_y({\bf y})}{\partial {\bf y}}\, d{\bf y}
= -{\bf I}_n$.
Thus, \eqref{eq:Cauchy-Schwarz inequality} gives
$({\bf v}^T{\bf I}_n{\bf v})^2
\le
({\bf v}^T{\bf I}_n{\bf v})
({\bf v}^T \e[\boldsymbol{\xi}({\bf y})\boldsymbol{\xi}^{T}({\bf y})]{\bf v})$,
which leads to 
${\bf I}_n \leq \e[\boldsymbol{\xi}({\bf y})\boldsymbol{\xi}^{T}({\bf y})]$.
Plugging this matrix inequality  into \eqref{eq:FIM theta1} yields
$\left(\frac{\partial{\bf m}}{\partial \boldsymbol{\theta}_1}\right)^{\!T}
{\bf \Sigma}^{-1}
\left(\frac{\partial{\bf m}}{\partial \boldsymbol{\theta}_1}\right)
\;\le\;
[{\bf I}(\boldsymbol{\theta})]_{(1,1)}$,
where the left-hand side is the $(1,1)$ block of the Gaussian FIM given by \eqref{eq:FIM Gaussian 11}.
Using \eqref{eq:CRB theta1 theta2}, we get ${\rm CRB}(\boldsymbol{\theta}_1)\le {\rm CRB}_{\rm Gau}(\boldsymbol{\theta}_1)$.
The equality in \eqref{eq:Cauchy-Schwarz inequality} occurs for 
${\bf v}^T{\bf y} p_y({\bf y})\propto
\frac{d p_y({\bf y})}{d{\bf y}}
{\bf v}, \forall {\bf v}\in  \mathbb{R}^{n}$, i.e. for $p_y({\bf y})=\frac{1}{(2\pi)^{n/2}}e^{-\|{\bf y}\|^2/2}$.
\hfill
\QED

This result constitutes an extension of the result reported in \cite{Stoica2011} in the univariate setting, where ${\theta}_1=m$.

Alternatively, identifying the distribution that minimizes the FIM with respect to $\boldsymbol{\theta}_2$ from expression \eqref{eq:score theta2} of the score $s_x(\boldsymbol{\theta}_2)$  appears to be highly challenging,  except in the case of a model with $n$ univariate i.i.d. r.v's, as demonstrated in Section \ref{sec:Elliptically symmetric distributions: covariance matrix parameterization}.
%
%%%%%%%%%%%%%%%%%%%%%%%%%%%%%%%%%%%%%%%%%%%%%%%%%%%%%%%%%%%%%%%%%%%%%%
\subsection{Impact of a non-symmetrical probability density function}
\label{sec:Impact of a non-symmetrical probability density function}
%%%%%%%%%%%%%%%%%%%%%%%%%%%%%%%%%%%
%
In this section, we present a counterexample showing that, for distributions whose p.d.f.'s are not symmetric about the mean, the Gaussian distribution does not necessarily maximize the CRB for the mean or the covariance parameter.
Consider $n$ independent Gamma-distributed observations ${\bf x}=(x_1,..,x_n)^T$, with joint p.d.f.
$p_x({\bf x})
=\prod_{k=1}^n \left(
\frac{1}{\Gamma(\alpha)\beta^{\alpha}}x_k^{\alpha-1}e^{-x_k/{\beta}}\mathds{1}_{[0,\infty)}(x_k)
\right)$
 where $\alpha>0$, $\beta>0$ and $\Gamma(\alpha)\pardef \int_0^{+\infty}t^{\alpha-1}e^{-t}dt$. The mean and the variance of $x_k$ are  $m=\alpha \beta$ and $\sigma^2=\alpha \beta^2$.
Using the one-to-one mapping $(\alpha,\beta) \leftrightarrow (m,\sigma^2)$,  $p_x({\bf x})$ can also be parameterized by $\boldsymbol{\theta}= (m,\sigma^2)^T$.
The expressions of the score functions 
$s_x(\alpha) = \sum_{k=1}^n (\ln x_k - \psi(\alpha) - \ln \beta)$
and
$s_x(\beta) = \sum_{k=1}^n (\frac{x_k}{\beta^2} - \frac{\alpha}{\beta})$
where $\psi(\alpha) \triangleq \frac{d}{d\alpha} \ln \Gamma(\alpha)$  is the digama function allows us to straightforward derive 
$\e[s_x^2(\alpha)] = n \psi'(\alpha)$, 
$\e[s_x^2(\beta)] = \frac{n \alpha}{\beta^2}$,
and $\e[s_x(\alpha) s_x(\beta)] = \frac{n}{\beta}$.
 Using the reparameterization in terms of $(m,\sigma^2)$, the following FIM is easily deduced:
 ${\bf I}(\boldsymbol{\theta})
 =n \tiny{\left(\begin{array}{cc}
\!\!\!\frac{4m^2}{\sigma^4}\psi'(\frac{m^2}{\sigma^2})-\frac{3}{\sigma^2}&-\frac{2m^3}{\sigma^6}\psi'(\frac{m^2}{\sigma^2})+\frac{2m}{\sigma^4}\!\!\!\\
\!\!\!-\frac{2m^3}{\sigma^6}\psi'(\frac{m^2}{\sigma^2})+\frac{2m}{\sigma^4}&\frac{m^4}{\sigma^8}\psi'(\frac{m^2}{\sigma^2})-\frac{m^2}{\sigma^6}\!\!\! \\
\end{array}
\right)}$.
From this, the inverse FIM provides the CRBs for $m$ and $\sigma^2$. 
Using the expansion:
$\psi'(u)=\frac{1}{u}+\frac{1}{2u^2}+\frac{1}{6u^3}+o(\frac{1}{u^3})$
\cite[rel. 6.4.12)]{Abramowitz1972}, we get the following result:
\begin{result}
\label{re:conter example CRB Gamma}
There exist non-symmetric p.d.f. about the mean for which the CRB on the mean equals that of the Gaussian distribution, while the CRB on the covariance exceeds that of the Gaussian distribution.
\begin{equation}
\label{eq:rel CRB Gamma Gaussian}
{\rm CRB}(m)
=\frac{\sigma^2}{n}={\rm CRB}_{\rm Gaus}(m)
\ \mbox{and}\
{\rm CRB}(\sigma^2)
=\frac{2\sigma^4}{n}\left(1+\frac{5}{3}\frac{\sigma^2}{m^2}+o(\frac{\sigma^2}{m^2})\right)>{\rm CRB}_{\rm Gaus}(\sigma^2)
\end{equation}
\end{result}
We note that for $\sigma \ll m$, equivalently  $\alpha \gg 1$, ${\rm CRB} (\sigma^2) \approx {\rm CRB}_{\rm Gaus}(\sigma^2)$. 
For integer $\alpha$, this is explained by the central limit theorem: the Gamma distribution is then the distribution of the sum of 
$\alpha$ independent exponential r.v.'s with parameter $\beta$.
%
%%%%%%%%%%%%%%%%%%%%%%%%%%%%%%%%%%%%%%%%%%%%%%%%%%%%%%%%%%%%%%%%%%%%%
\subsection{Elliptically symmetric distributions: covariance matrix parameterization}
\label{sec:Elliptically symmetric distributions: covariance matrix parameterization}
%%%%%%%%%%%%%%%%%%%%%%%%%%%%%%%%%%%
%
In this section, we consider the important case where the p.d.f. of the normalized random vector r.v.
${\bf y}\pardef {\bf \Sigma}^{-1/2}({\bf x}-{\bf m})$
is not only symmetrical about the mean  but also spherically symmetrical, i.e., $p_y({\bf y})=g(\|{\bf y}\|^2)$, where $g(.)$ satisfies the normalizing constraint 
$\int_0^{\infty}t^{n/2-1}g(t)dt=\delta_n$ with $\delta_n= \Gamma(n/2)/ \pi^{n/2}$
and the constraint $\cov({\bf y})={\bf I}_n$. This co-called density generator $g(.)$ may sometimes depend on a nuisance parameter $\boldsymbol{\theta}_3$.
This leads to the family of elliptically symmetric distributions with  p.d.f.:
\begin{equation}
\label{eq:pdf ES}
 p_x({\bf x})= |{\bf \Sigma}|^{-1/2}g \left(({\bf x}-{\bf m})^T{\bf \Sigma}^{-1}({\bf x}-{\bf m})\right)\!.
\end{equation}
The choice of $g(.)$ distinguishes different families of elliptically symmetric distributions, among which the Student-$t$ and the generalized Gaussian distributions are widely used \cite{Abeida2023}. This class of distributions has several useful properties. In particular, the r.v. ${\bf x}$ admits the stochastic representation\footnote{where $x=_d y$ means that the r.v. $x$ and $y$ have the same distribution.} \cite{Delmas2024}:
\begin{equation}
\label{eq:SR}
{\bf x}
=_d {\bf m}+ \sqrt{\mathcal{Q}}{\boldsymbol{\Sigma}}^{1/2} {\bf u},
\end{equation}
where $\mathcal{Q}$ is a positive r.v.,   $ {\bf u}$ is uniformly distributed on the unit $n$-sphere, and $\mathcal{Q}$  and $ {\bf u}$ are independent.
The constraint $\cov({\bf y})={\bf I}_n$ imposes $\e(\mathcal{Q})=n$, and \eqref{eq:SR} implies 
$\mathcal{Q}=_d ({\bf x}-{\bf m})^T{\bf \Sigma}^{-1}({\bf x}-{\bf m})$ with p.d.f.
\begin{equation}
\label{eq:pdf q}
p(q)= \delta_n^{-1}q^{n/2-1}g(q).
\end{equation}
This family of distributions has been widely used in various engineering applications requiring non-Gaussian models (see \cite{Delmas2024} for further details).

In the absence of nuisance parameter $\boldsymbol{\theta}_3$, the FIM for $\boldsymbol{\theta}_1$ and $\boldsymbol{\theta}_2$ are given by the Slepian-Bang formula \cite{Besson2013}, which generalizes the Gaussian FIM \eqref{eq:FIM Gaussian 11},\eqref{eq:FIM Gaussian 22}
\begin{align}
\label{eq:FIM ES11}
[{\bf I}(\boldsymbol{\theta})]_{(1,1)}
&=
a_{0,n}(g)\left(\frac{d{\bf m}}{d\boldsymbol{\theta}_1}\right)^T{\bf \Sigma}^{-1}\frac{d{\bf m}}{d\boldsymbol{\theta}_1}
\\
\nonumber
[{\bf I}(\boldsymbol{\theta})]_{(2,2)}
&=
a_{1,n}(g)
\left(\frac{d \ve({\bf \Sigma})}{d\boldsymbol{\theta}_2}\right)^T({\bf \Sigma}^{-1}\otimes {\bf \Sigma}^{-1})
\frac{d \ve ({\bf \Sigma})}{d\boldsymbol{\theta}_2}
\\
\label{eq:FIM ES22}
&+
a_{2,n}(g)
\left(\frac{d \ve({\bf \Sigma})}{d\boldsymbol{\theta}_2}\right)^T \ve({\bf \Sigma}^{-1})\ve^T({\bf \Sigma}^{-1})
\frac{d \ve ({\bf \Sigma})}{d\boldsymbol{\theta}_2},
\end{align}
where the coefficients $a_{0,n}(g)$, $a_{1,n}(g)$ and  $a_{2,n}(g)$ are $g$-dependent and are given by:
$a_{0,n}(g)
=\frac{\e[\mathcal{Q}\varphi^2(\mathcal{Q})]}{n}$, 
$a_{1,n}(g)
=\frac{1}{2}\frac{\e[\mathcal{Q}^2\varphi^2(\mathcal{Q})]}{n(n+2)}$ 
and  
$a_{2,n}(g)
=\frac{1}{4}(2a_{1,n}(g)-1)$ with $\varphi(t)\pardef  -\frac{2}{g(t)}\frac{d g(t)}{dt}$. 
Comparing \eqref{eq:FIM ES11},\eqref{eq:FIM ES22} to  the Gaussian case \eqref{eq:FIM Gaussian 11},\eqref{eq:FIM Gaussian 22}, we have
$(a_{0,n}(g),a_{1,n}(g),a_{2,n}(g))=(1,1/2,0)$ for a Gaussian distribution.

Consistent with Result \ref{re:Gaussian best}, it is proven in \cite{Abeida2019} that $a_{0,n}(g)\ge 1$ and $a_{0,n}(g)= 1$ i.i.f. 
$g(t)=(2\pi)^{-n/2}e^{-t/2}$, i.e., when ${\bf x}$ is Gaussian distributed.

We now consider the case in which the parameter of interest is $\boldsymbol{\theta}_2$.  the following result is proved.
\begin{result}
\label{re:conter example RES}
Among elliptically symmetric distributions in which the parameter of interest is embedded in the covariance matrix ${\bf \Sigma}$, the following statements hold:
\begin{enumerate}
\item[(i)]
There exists an upper bound on the CRB associated with $\boldsymbol{\theta}_2$. This upper bound is asymptotically attained by the generalized Gaussian distribution\footnote{with density generator $g(t)= b_{n,s}e^{-c_{n,s}t^s}$, where $b_{n,s}$ and $c_{n,s}$ are given in \cite{Abeida2023}.} as the shape (exponent) parameter $s$ approaches zero.
\item[(ii)]
For a model consisting of $n$ i.i.d. observations ${\bf x}=(x_1,\ldots,x_n)^T$, the generalized Gaussian distribution asymptotically maximizes the CRB on $\boldsymbol{\theta}_2$ among all distributions  that are symmetric about the mean. In this case,
${\rm CRB}(\sigma^2)=\frac{1}{n}\frac{2\sigma^4}{s}$, which is unbounded as $s \to 0$.
\item[(iii)]
Within the subclass of compound Gaussian distributions, the CRB on $\boldsymbol{\theta}_2$ is always larger than that obtained for the Gaussian distribution.
\end{enumerate}
\end{result}
{\it Proof:}
Applying the Cauchy-Schwarz inequality, we get
$[\e(\mathcal{Q}\varphi(\mathcal{Q})) ]^2=[\e(1.(\mathcal{Q}\varphi(\mathcal{Q}))) ]^2 \le \e(1^2) \e(\mathcal{Q}^2\varphi^2(\mathcal{Q}))$ 
where from \eqref{eq:pdf q},
$\e(\mathcal{Q}\varphi(\mathcal{Q}))
=-2\delta_n^{-1}\int_0^{\infty}q^{n/2}dg
=-2\delta_n^{-1}[q^{n/2}g(q)]_{q=0}^{q=\infty}+n\delta_n^{-1}\int_0^{\infty}q^{n/2-1}g(q)dq=n$ because 
$\e(\mathcal{Q})=\delta_n^{-1}\int_0^{\infty}q^{n/2}g(q)dq=n$ implies $\lim_{q \rightarrow \infty}q^{n/2}g(q)=0$.
So we get $n^2 \le 1  \times 2 a_{1,n}(g) n(n+2)$.
Consequently $a_{1,n}(g) \ge \frac{n}{2(n+2)}$ and $a_{2,n}(g) \ge -\frac{1}{2(n+2)}$
and these two lower bounds are both lower than those of the Gaussian distribution.
These two lower bounds are not reached because there is no constraint  function $g(.)$ such that $ g(t)\propto t \frac{dg}{dt}$.
But the coefficients $(a_{1,n},a_{2,n})$ associated with the generalized Gaussian distribution of exponent parameter $s>0$, which controls the non-Gaussianity are given by \cite{Abeida2019} $(a_{1,n},a_{2,n})=(\frac{n+2s}{2(n+2)},-\frac{1-s}{2(n+2)})$ when the parameter $s$ is known. The proof of (i) follows because 
$\lim_{s \rightarrow 0}(a_{1,n},a_{2,n})=(\frac{n}{2(n+2)},-\frac{1}{2(n+2)})$ and because 
$[{\bf I}(\boldsymbol{\theta})]_{(2,2)}$ given by \eqref{eq:FIM ES22} is a linear combination of two definite symmetric matrices.
\hfill
\QED

For the model of $n$ independent generalized Gaussian distributed r.v., the FIM of a single r.v. $x_k$ is given by \eqref{eq:FIM ES22} for $n=1$ with
 $(a_{1,1},a_{2,1})=(\frac{1+2s}{6},-\frac{1-s}{6})$ and consequently ${\rm CRB}(\sigma^2)=\frac{1}{n}\frac{2\sigma^4}{s}$ proves (ii).
\hfill
\QED

The compound Gaussian distributions are characterized by a density generator of the form
$g(t)=(2\pi)^{-n/2}\int_0^{\infty}\tau^{-n/2}e^{-t/2\tau}dF_{\tau}(\tau)$ where $\tau$ is the so-called texture.
Using the relation \cite{Abeida2020}
$\frac{1}{2}\frac{\e[\mathcal{Q}^2\varphi^2(\mathcal{Q})]}{n(n+2)}=\frac{1}{2}
+\frac{\e[\mathcal{Q}^2{\varphi}'(\mathcal{Q})]}{n(n+2)}$ 
with
${\varphi}'(t)=-2\frac{(g^{''}(t)g(t)-[g^{'}(t)]^2)}{g^2(t)}$ where
$g^{(k)}(t)=(-2)^{-k}(2\pi)^{-n/2}\int_0^{\infty}\tau^{-n/2-k}e^{-t/2\tau}dF_{\tau}(\tau)$,
$k=0,1,2$,
we get
$a_{1,n}(g)
=\frac{1}{2}+\frac{\e[\mathcal{Q}^2{\varphi}'(\mathcal{Q})]}{n(n+2)}$. 
Since ${\varphi}'(t) \le 0$ i.f.f.
$g^{''}(t)g(t) \ge[g^{'}(t)]^2$ which follows from the Cauchy-Schwarz inequality, we have
$a_{1,n}(g) \le \frac{1}{2}$ and $a_{2,n}(g) \le 0$,
with equality i.f.f. $\tau=1$, i.e., $g(t) = (2\pi)^{-n/2} e^{-t/2}$, so that ${\bf x}$ is Gaussian distributed, which proves (iii).
\hfill
\QED

Finally, in the presence of nuisance parameter $\boldsymbol{\theta}_3$ (e.g., for $s$ unknown for the generalized Gaussian distribution), it is proved in \cite{Abeida2023} that $[{\bf I}(\boldsymbol{\theta})]_{(2,2)}-[{\bf I}(\boldsymbol{\theta})]_{(2,3)}[{\bf I}(\boldsymbol{\theta})]_{(3,3)}^{-1}[{\bf I}(\boldsymbol{\theta})]_{(2,3)}^T$ whose inverse is the CRB on $\boldsymbol{\theta}_2$ is also given by \eqref{eq:FIM ES22} where the coefficient 
$a_{2,n}(g)$ is replaced by $a_{2,n}(g)+a_{3,n}(g)$ with $a_{3,n}(g)<0$.
It follows that ${\rm CRB}(\boldsymbol{\theta}_2)$ will always be greater than in the absence of nuisance parameters.
 This is illustrated in Fig. 1 for 
$n$ independent generalized Gaussian distributed observations ${\bf x}=(x_1,..,x_n)^T$, whose p.d.f. is 
$p_x({\bf x})
=\prod_{k=1}^n \left(
\frac{b_{1,s}}{\sigma}
e^{-c_{1,s}(\frac{x_k-m}{\sigma})^{2s}}
\right)$ for $s>0$, where ${\rm CRB}(\sigma^2)\!>\!{\rm CRB}_{\rm Gaus}(\sigma^2)$ for $s\!<\!1$, 
${\rm CRB}(\sigma^2)\!<\!{\rm CRB}_{\rm Gaus}(\sigma^2)$ for $s\!>\!1$ and
${\rm CRB}(\sigma^2)\!=\!{\rm CRB}_{\rm Gaus}(\sigma^2)=2\sigma^4/n$ for $s\!=\!1$ (Gaussian distribution).

\vspace{-4.5cm}

%%%%%%%%%%%%%%%%%%%%%%%%%%%%%%%%%
%%%%%%%%%%%%%%%%%%%%%%%%%%%%%%%%%
\begin{figure}[H]
\begin{center}
{\psfig{file=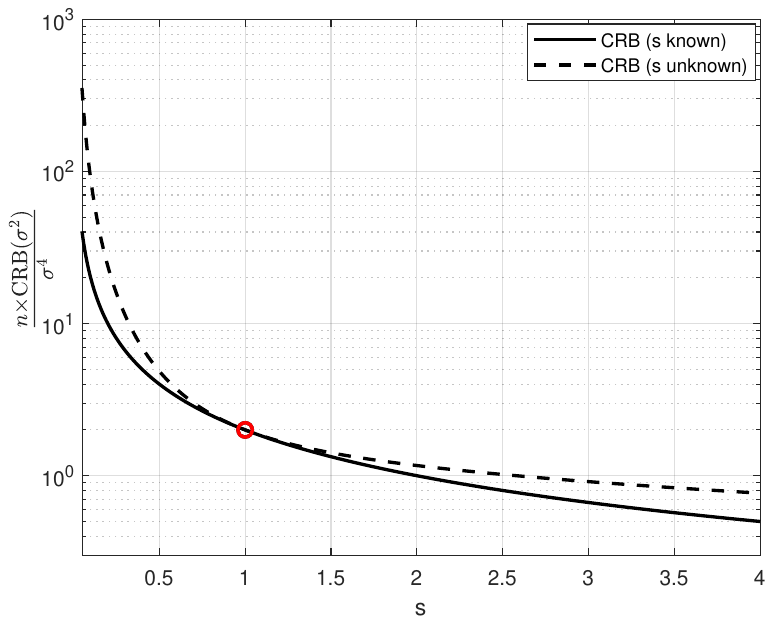,width=6in,height=5.5in}}
\end{center}

\vspace{-5cm}

\begin{center}
{\small \footnotesize
Fig.1.
$n{\rm CRB}(\sigma^2)/\sigma^4$ as a function of the exponent parameter $s>0$.}
\end{center}
\end{figure}
%
%%%%%%%%%%%%%%%%%%%%%%%%%%%%%%%%%%%%%%%%%%%%%%%%%%%%%%%%%%%%%%%%%%%%%%
\section{What we have learned}
\label{sec:What we have learned}
%%%%%%%%%%%%%%%%%%%%%%%%%%%%%%%%%%%
%
The results presented in the previous sections indicate that {\it the Gaussian data assumption leads to the largest CRB under only three conditions}:
(i) the p.d.f. of the data ${\bf x}$ is symmetric about the mean  (ensuring the decoupling between the parameters of the mean and those of the covariance in the FIM),
(ii) the parameter of interest belongs to the mean; and
 (iii) there is no additive nuisance parameter.
 If any of these conditions is not satisfied, no general result exists regarding the distribution that maximizes the CRB on the parameter of interest.
 In particular, a counterexample with a non-symmetric distribution shows that the CRB on the mean equals that of the Gaussian distribution, while the CRB on the covariance exceeds the Gaussian CRB. Moreover, examples from the class of elliptically symmetric distributions demonstrate that, even for distributions symmetric about the mean, the CRB on parameters of the covariance matrix can be either smaller or larger than the corresponding Gaussian CRB. Within the subclass of compound Gaussian distributions, however, the CRB for covariance parameters is always larger than that of the Gaussian.
%

%%%%%%%%%%%%%%%%%%%%%%%%%%%%%%%%%%%%%%%%%%%%%%%%%%%%%%%%%%%%%%%%%%%%%%
\section{Acknowledgement}
\label{sec:Acknowledgement}
Prof. Habti Abeida would like to acknowledge the Deanship of Graduate Studies and Scientific Research, Taif University, for funding this work.

%%%%%%%%%%%%%%%%%%%%%%%%%%%%%%%%%%%%%%%%%%%%%%%%%%%%%%%%%%%%%%%%%%%%%%
\section{Authors}
\label{sec:Authors}
%%%%%%%%%%%%%%%%%%%%%%%%%%%%%%%%%%%
%
{\bf Jean-Pierre Delmas} (jean-pierre.delmas@it-sudparis.eu) received the French HDR degree from the University of Paris XI, Orsay, in 2001.
He is currently a Professor with the Samovar laboratory, Telecom SudParis, Institut Polytechnique de Paris, Palaiseau, France.
His research interests lie in statistical methods for signal processing with emphasis on asymptotic performances analysis. 

{\bf Habti Abeida} (h.abeida@tu.edu.sa) received the Ph.D. degree in signal processing and applied mathematics in 2006 from Pierre and Marie Curie University, Paris, France. He is currently a Professor with the Electrical Engineering Department at Taif University, Taif, Saudi Arabia. His research interests include statistical signal processing with applications to radar, antenna arrays, and communications.
%
%%%%%%%%%%%%%%%%%%%%%%%%%%%%%%%%%%%%%%%%%%
%%%%%%%%%%%%%%%%%%%%%%%%%%%%%%%%%%%%

%%%%%%%%%%%%%%%%%%%%%%%%%%%%%%%%%%%%%%%
\end{document}